\documentclass[preprint, aps,prl,groupedaddress]{revtex4} 
\pdfoutput=1
\usepackage{graphicx}% Include figure files
\usepackage{upgreek} 
\usepackage{amssymb}      
\usepackage{amsmath}  
\usepackage{textpos}  
\usepackage{hyperref}
\hypersetup{colorlinks=true, citecolor=blue, urlcolor=blue, linkcolor=blue} 

\begin{document}  
\begin{textblock*}{\textwidth}(0cm,0cm)
{{\it This is the accepted author version of the manuscript published in} Physical Review Applied vol. 7, 034001 (2017) [$\copyright$ 2017 American Physical Society]. The version of record is available from the publisher's website from the link: \url{https://doi.org/10.1103/PhysRevApplied.7.034001}}
\end{textblock*}  
\vspace{4cm} 
\title{Broadband near-unidirectional absorption enabled by phonon-polariton resonances in SiC micropyramid arrays}        
\author{G. C. R. Devarapu$^{1,2,3}$ and S. Foteinopoulou*$^{1,4}$} 
\address{$^{1}$ School of Physics, College of Engineering, Mathematics and Physical Sciences (CEMPS),University of Exeter, Exeter, EX4 4QL, United Kingdom.}
\address{$^{2}$ School of Physics and Astronomy, University of St. Andrews, North Haugh, St Andrews, KY16 9SS, United Kingdom.}
\address{$^{3}$ Cork Institute of Technology, Bishoptwon, Cork, T12 P928, Ireland.}
\address{$^{4}$ Center for High Technology Materials, University of New Mexico, 1313 Goddard St. SE, Albuquerque, New Mexico 87106, USA.}
\begin{abstract}        
{Inspired by moth eyes, nature's most powerful antireflex, we present a sub-wavelength SiC micropyramid design, which operates in the Reststrahlen band of SiC, namely the spectral band of strong phonon-photon coupling in the SiC material. While within this band SiC repels EM waves, we observe here a broad low-reflectivity window with unique attributes, with distinct characteristics different from typical dielectric moth-eye-like structures. To be specific, while the latter systems are entirely symmetric, the reflection response of our SiC micropyramid system can be highly asymmetric. In particular, the SiC micropyramid system can be near-reflectionless for light impinging from the tip side of the micropyramids and exhibits more than 90$\%$ reflection for light impinging from the base side of the micropyramids, over a broad wavelength range in the SiC Reststrahlen band.  This strongly asymmetric reflection response emanates from the cascaded coupling of vortex-like cavity modes at each of the SiC blocks comprising the micropyramids and translates into a strongly uni-directional absorber response. We discuss how, by virtue of Kirchhoff's law, this strongly uni-directional superabsorber behavior implies a strongly uni-directional emission profile that is important for one-way infrared sources and passive radiative-cooling systems. }      
\end{abstract}
     
\maketitle          
%--------------------------
\section{I. INTRODUCTION}
%--------------------------
\par
Across the EM spectrum, recent intense research efforts aim to uncover new mechanisms for extreme absorption control that go beyond the traditional three-component platforms with the absorber sandwiched between an anti-reflex coating and a backreflector. The driving force behind these investigations are  photovoltaic devices \cite{atwater, aydin, brongersma, povinell}  as well as infrared/THz detectors \cite{padilla1, detec1, wasser1, detec2, kats1}, modulators \cite{modul, kats1} and thermal-emitters \cite{wasser1, wasser2, shvets1, kats1}. The strategies underpinning these current absorption management efforts could be grouped into five general approaches. In particular, new superabsorber architectures that have been proposed may utilize plasmonic resonances that enhance the EM field in the vicinity of the absorber  \cite{atwater, aydin, durdu1}, involve photonic metamaterials \cite{shalaevbook, soukreview} either as impedance matchers \cite{padilla1, padilla2, metaabsor} or enhancers of dark EM field components by virtue of a characteristic hyperbolic photonic dispersion  \cite{boltasseva}, employ lossy material bilayers with destructive Fabry-Perot  interference suppressing reflection \cite{kats1, kats2}, tailor resonant coupling to waveguide/cavity modes in the absorber structure \cite{povinell, shvets1}, or manipulate a near-reflectionless coupling to the near band-edge Floquet-Bloch mode of a lossy photonic crystal \cite{sf1, sf2}. 
\par
Of the aforementioned avenues to extra-ordinary absorption management the last two are particularly attractive as they essentially represent one-step super-absorber platforms. That is to say the functional performance of these structures relies only on a single kind of absorbing medium; it is not needed to incorporate other type of metallic or absorber material that may make fabrication more complicated, lead to parasitic absorption \cite{brongersma, padilla1} or compromise structural stability due to overheating. This is not the case for any of the other schemes, where together with the absorber of interest, additional patterned absorber/metallic material are integrated to enable impedance-matching \cite{photobook1, padilla1, kats2}, EM-field enhancement \cite{atwater, brongersma}, or additional light passes \cite{photobook1, biswas}.    
\par
Now, the valuable insight obtained for absorption with plasmonic nanostructures in the visible spectrum \cite{aydin} can be transferred in the mid-IR spectral regime, albeit not with the use of metals. Metals in the infrared spectrum have a very small skin depth; thus they repel the EM field and do not respond with the familiar plasmonic resonances existing in the visible spectrum. However, ``plasmonic effects'' in the mid-IR can be mimicked if we instead consider microstructures comprising material with a strong phonon-photon coupling. Such phonon-polariton materials can exhibit a negative permittivity response in the mid-IR with a larger skin depth. They so emulate field-confinement effects with infrared light similar to those of plasmonic nanostructures \cite{hessel_rev} operating in the visible spectrum. 
\par
Thus, for mid-IR absorption platforms, SiC would be an excellent material with its phonon-polariton gap spectrum (Reststrahlen band) between 10.3 to 12.6 $\mu$m \cite{catrysse}. This spectral regime is of high interest for detectors and sources as it encompasses the absorption fingerprints of many biomolecules \cite{biosens}. Moreover, this spectrum is also highly relevant to passive radiative-cooling devices \cite{fan_rad}, since it lies within the atmospheric transparency window \cite{atmosph, atmo}. In addition, SiC is a refractory material, which is crucially important for the structural stability of absorber/emitter devices \cite{shalaev_refra}.  
\par
In this paper, we investigate SiC-based designs with the objective to achieve mid-IR sub-wavelength broadband superabsorption\cite{superabsordef, aydin, superabsordef2,superabsordef3} that is also highly uni-directional. Designing and controlling a  highly unidirectional mid-IR absorption response, has not been thus-far investigated but can benefit various applications. In particular, highly unidirectional absorbers would efficiently absorb light that impinges from their top side while they would reflect, and thus not absorb, almost all light, that impinges from their bottom side. Kirchhoff's law then prescribes \cite{Kirchhoff1, Kirchhoff2}, that these systems would emit light predominantly towards the direction where they are strong absorbers  if light was incident from that direction, in this manner acting as near-unidirectional emitters. Uni-directional emission is a highly attractive feature not only for the design of one-way infrared sources but also for simpler integrated components for passive-radiative-cooling devices to reduce temperature in buildings \cite{fan_rad} or electronic circuits. In order to achieve this goal we will bring together two of the aforementioned explored avenues for harnessing absorption, i.e. invoking cavity resonances and photonic crystal (PC) effects and combine them with intuition obtained from the anti-reflex properties of moth-eye-like structures. This allows to consider SiC-only platforms without other auxiliary absorbing material structures that could compromise the resilience of the device under high temperatures.                 
\par  
This paper is organized as follows: In Sec. II we present the SiC-based platform we envision for strong absorption control within the Reststrahlen band of SiC. In Sec. III we present our results for an initial design demonstrating enhanced broadband absorption. We discuss the origin of such absorption enhancement with respect to bulk SiC in the same frequency region in Sec. IV. In Sec. V we analyze the effects of a stronger interaction between the micropyramid building-blocks towards a broadband superabsorber\cite{aydin, superabsordef, superabsordef2, superabsordef3} behavior. In Sec. VI we quantify the absorption enhancement and demonstrate the robustness under misalignment of the micropyramid superabsorber platform. In Sec. VII we investigate the asymmetric absorber response for light illuminating from the tip-side and base-side of the micropyramid-array platform. In Sec. VIII we discuss how this asymmetry can be harnessed further towards a near-unidirectional absorber/emitter behavior. Finally, In Sec. IX we present our conclusion. 

\section{II. THE Silicon Carbide PLATFORM FOR BROADBAND SUPERABSORPTION}  
\par
In spite of Silicon Carbide's high loss tangent within its Reststrahlen band, which makes it potentially attractive for absorber/emitter systems, it is actually extremely difficult to couple light inside the SiC material in this spectrum; most light gets reflected upon hitting the SiC surface. This low in-coupling poses a serious bottleneck in SiC-based absorber/emitter platforms. However, recent results with a compact one-dimensional SiC judicious PC design \cite{sf2} show that it is possible to suppress such reflection and achieve near-perfect absorption within a narrow spectral band. This band can be extended to cover most of the SiC Reststrahlen spectral region, by employing a very thick, several wavelengths long, PC \cite{jeosrp}. However, in practical applications it is desirable to have compact designs exhibiting a broadband  super-absorption \cite{superabsordef, superabsordef2, superabsordef3}.  In addition, we seek such super-absorber merit to be highly asymmetric, i.e. to exist only for light impinging from one side of the structure, a possibility that has not been hitherto explored. 
\par
In order to achieve our target of a broadband absorption in SiC within its Reststrahlen band with a compact structure, we employ a design with a SiC micropyramid  as its elementary building block arranged in a one-dimensional photonic crystal array. We draw our inspiration  for such a design from the conical shape of corneal nipples in moth eyes \cite{moth1, moth2}, which are known for their superior antireflection properties.  The moth-eye corneal nipples do not constitute an optically dense material. So, naturally one may wonder whether designs of a similar shape  but made from a material with extreme optical properties, such as SiC in the Reststrahlen band, would retain these extra-ordinary antireflex properties.  Antireflection behavior has been demonstrated \cite{rivas, biomime2} in the optical spectrum with nanostructured cones or pyramids mimicking moth eyes comprising a high-refractive-index material, such as GaP or Si, which have a purely dielectric optical response. These results provide a promising basis to expand such antireflex behavior in a platform made of SiC, whose material optical response in the Reststrahlen band involves both a negative permittivity and an optical loss tangent (see bottom-left inset in Fig. 1), due to the photon-phonon coupling resonance in this spectrum. Thus we proceed with our intuitive choice of SiC micropyramid-based designs which we investigate further in the following. 
  
\par  
We show the design under consideration in Fig. 1. We consider a stepped two-dimensional (2D) pyramid, with translational symmetry in the third dimension. The translational symmetry ensures full polarization decoupling \cite{pwe}, allowing for polarization selective emission suitable for constructing polarized infrared sources. Moreover, we are considering a stepped micropyramid structure, whose realization is possible with certain fabrication techniques, such as direct wafer-bonding \cite{SiCwafer} or layer-by-layer fabrication with e-beam patterning \cite{gsubra_ebeam} and inductively coupled plasma reactive-ion etching (ICP-RIE) \cite{SiCetch}.  
\par
Each block of the micropyramid is expected to behave as a resonant cavity \cite{bronger2} microantenna with emission/absorption spectra in the mid-IR regime. By combining blocks of different sizes we aim to bring together different resonant frequencies, thus constructing a broadband response for the pyramid microantenna. Then by arranging the pyramid blocks in a moth-eye structure we aim to minimize reflection thus enabling a strong EM in-coupling to each of the participant block microresonators. That is to say that our proposed design is conceived with a vision to operate simultaneously as a moth-eye antireflector and a broadband micro-antenna.  
\par
The sizes of the participating SiC blocks in the pyramid of Fig. 1 are chosen so that a gradual size variation from one block to the next one is maintained, with a small number of layers -which is desirable from the fabrication point of view-. At the same time, we aim for a compact structure of sub-wavelength total thickness, i.e. thickness in the order of half the free space wavelength. In Fig. 1 we label each of the blocks comprising the micropyramid with 1, 2, 3 and 4. Their corresponding sizes (in the {\it x}-- and {\it y}-- direction respectively) are: $\textrm w_{\textrm x1} = 0.5 \hspace{1mm} \mu \textrm m$, $\textrm w_{\textrm y1} = 1 \hspace{1mm} \mu \textrm m$,  $\textrm w_{\textrm x2} = 1 \hspace{1mm} \mu \textrm m$, $\textrm w_{\textrm y2} = 1 \hspace{1mm} \mu \textrm m$,  $\textrm w_{\textrm x3} = 2 \hspace{1mm} \mu \textrm m$, $\textrm w_{\textrm y3} = 1 \hspace{1mm} \mu \textrm m$, $\textrm w_{\textrm x4} = 4 \hspace{1mm} \mu \textrm m$, $\textrm w_{\textrm y4} = 1 \hspace{1mm} \mu \textrm m$. 
\par 
These micropyramids are arranged periodically in an array with a spacing, $\textrm  a$. We will vary this inter-spacing, $\textrm  a$, so we can control the system's collective response from a regime where each micropyramid responds independently to the impinging light to a regime where the resonances between neighboring micropyramids are interacting.  In our initial investigation, we will ignore the presence of a required substrate, seen in Fig. 1, in order to capture and understand the response of our proposed microstructure better. We will then consider the effect of the substrate afterward. So, unless a presence of a substrate is explicitly stated, all results correspond to the platform of Fig. 1, but without the depicted substrate.      
\par  
\section{III. ENHANCED ABSORPTION WITH THE SiC-MICROPYRAMID SYSTEM}  
\par
We first study the response of the micropyramid-array system, where each micro-pyramid building block would respond independently to the impinging light. For this purpose the micropyramids are placed sparsely, about one free-space wavelength apart, with an inter-spacing of $\textrm a= 10 \hspace{1mm} \mu$m. We consider normally incident light and both polarization cases, where the impinging light has its electric field aligned with the micropyramid block-axis $\it z$ (TE-polarization) or its magnetic field aligned with the micropyramid block-axis $\it z$ (TH-polarization). We use a two-dimensional implementation of the Finite-Difference Time-Domain (FDTD) method with the Lumerical FDTD simulator \cite{lumerical} to investigate the response of the micropyramid-array system. 
\par  
       
\par
In the FDTD simulator, periodic boundary conditions are applied in the $\it x$-direction as the micropyramid building block repeat themselves periodically in this direction. Open (absorbing) boundary conditions are applied in the $\it y$-direction to emulate an unbounded domain above and below the micropyramid. The micropyramid structure is discretized in the simulation with a mesh size of $dx= dy=10$ nm. In our calculations, we consider a SiC permittivity function characteristic of the phonon-polariton resonant response in the Reststrahlen band, i.e.:
\begin{equation}  
\varepsilon(\omega)=\varepsilon_{\infty}(1+\frac{\omega_L^2-\omega_T^2}{\omega_T^2-\omega^2-i\omega\Gamma})
\end{equation}    
The high frequency response, $\varepsilon_{\infty}$, the longitudinal and transverse optical phonon frequencies, $\omega_L$ and $\omega_T$, respectively, and the intrinsic damping parameter $\Gamma$ are determined by fitting to experimental optical data \cite{catrysse}. From Ref. \onlinecite{catrysse} we take: \hspace{1mm} $\varepsilon_{\infty}=6.7$, $\omega_L=2\pi\times 29.07$ THz, $\omega_T=2\pi\times 23.79$ THz, and $\Gamma=2\pi\times 0.1428$ THz. The SiC permittivity spectral function corresponding to Eq. 1 is depicted in the bottom-left inset of Fig. 1.
\par
We clearly observe in Fig. 2 that the micropyramid-array system responds very differently to TE and TH light. For TE light, the system shows both very little reflection and very little absorption [solid-black lines of panel (a) and (b) respectively]. This is because the electric field drives the phonon-polariton resonance along the $z$--direction, where the SiC material is unbounded. Thus, the EM field behaves as in the vicinity of bulk SiC, and only penetrates very little the SiC micropyramid. However, since the micropyramid array is arranged sparsely the EM field can flow around the structure yielding a very low absorption and high transparency. Indeed, we confirmed that as we bring the micropyramids closer the reflective response of the micropyramid system to TE light becomes similar to that of bulk SiC. Clearly, the SiC-micropyramid array structure is a weak absorber/emitter for TE-polarized light. 
\par
On the other hand, for TH-polarized light we observe in Fig. 2 a wide low-reflectivity window within the SiC Reststrahlen band between approximately 10.3 and 11.1 $\mu$m [dashed-red line in (a)], in spite of bulk SiC being nearly-perfectly reflecting in this range [dotted-green line in Fig. 2(a)]. This low-reflectivity window translates into a broad absorption window with absorption exceeding $50\%$ for the most part and reaching values as high as $80\%$ [dashed-red line in Fig. 2(b)]. This highly efficient absorption represents extraordinary absorption-enhancement factors, as high as 16, with respect to the absorption capabilities of bulk SiC [dotted-green line in Fig. 2(b)]. We note that a second window that we observe with relatively low reflection, appearing close to the red bound of the Reststrahlen band, is not accompanied by a high absorption. Clearly, we see in Fig. 2(b) that absorption is negligible within that range. We note in passing, that these two aforementioned spectral windows are separated by a spectrum with a high reflection response. In this intermittent regime, reflection reaches a maximum of about $80\%$  at an impinging wavelength of 11.27$\mu$m, with an asymmetric peak that is typical of Fano resonances \cite{fanoyuri}. 

\par
\section{IV. UNDERSTANDING THE ABSORBER BEHAVIOR OF THE SIC MICROPYRAMID SYSTEM} 
\par

Clearly, the results in Fig. 2 demonstrate that the design of Fig. 1 has a strong potential as a super-absorber for TH-polarized light, which we will explore further in the following. Therefore, we attempt to understand better the response of the micropyramid array by looking at the spectral response of each element it comprises. In other words, we will look at the spectral response of a periodic array comprised only of the top block, the second block, the third block and the fourth block of the micropyramid respectively, as designated in the bottom panel of Fig. 1. We show these results for the reflection, R, and absorption A, in Figs. 3(a) and 3(b) respectively, with a dotted-red line (top-block array), solid-black line (second-block array), dashed-green line (third-block array) and dot-dashed-blue line (fourth-block array). The arrays comprising either of the first two blocks have near-zero reflective properties. These block sizes are deep sub-wavelength, thus only minimally disturb the path of the incident light. However, clear resonances still exist as we can observe by the peaks in the absorption spectrum and the enhanced electric field within the SiC block. For example, see the inset depicting the electric field intensity enhancement within the first block for the free space wavelength M1. 
\par
The behavior is similar for the arrays comprised of the either the third or fourth block, as we can see from the peaks in the absorption spectrum and the electric field enhancement inside the blocks (for example see the inset for the fourth block for the free space wavelength M1). All blocks show a resonant response over a relatively broad spectrum, with the larger blocks sustaining resonances through longer wavelengths. We observe in Fig. 3 that for the large blocks the resonances are characterized by strong fields at the corners while for the smaller block the field is more uniform, with stronger fields at the sides of the block.  
       
\par 
The resonances and their corresponding electric field morphology arises from the surface phonon-polaritons at the facets of the SiC blocks and their interaction. Note that it is the strong phonon-photon coupling in SiC that yields a negative permittivity to SiC, thus enabling surface bound modes similar to those of metals at optical frequencies \cite{Wassermanreview}. As a result of such surface bound modes, resonant optical trapping occurs within the blocks as one can observe from the EM energy circulation plots (white steamlines) in panels (c) and (d) of Fig. 3 for the cases of the first and fourth block respectively. We can clearly identify vortices in the EM circulation, in either the sides [case seen in Fig. 3(c)] or in the vicinity of the corners [case seen in Fig. 3(d)]. These vortices signify the existence of cavity-like optical trapping behavior \cite{vortexcavity}.  On the other hand, for the lower frequencies near the red bound of the Reststrahlen band, we do not find any block cavity modes, as the field goes around the block [for example see the insets in Fig. 3 for blocks 1 and 4 for frequency M3]. 
\par
It is interesting to observe that the arrays comprising the third or fourth block individually demonstrate a clear asymmetric reflection peak as we see in Fig. 3(a), which separates the spectral regime where localized resonant modes exist within the block, --as for example we see from the electric field inset for the fourth block at free space wavelength M1, from the spectral regime where void modes are excited in the interstitial region between the neighboring blocks, -- as for example we see from the electric field inset for the fourth block at free space wavelength M3--. In these void-modes the electric field is near-zero within the block. 
\par
It is the former spectral regime, i.e. the regime of block-cavity resonances, that is evidently of interest with respect to achieving superabsorption. This is because in order to harness a strong absorption response, the strong electric fields must spatially overlap with the absorbing matter. This follows directly from Poynting's theorem for power dissipation [ see also eq. A3]. In passing, we also note that the presence of the spectrally asymmetric reflection peak observed in Fig. 3, suggests the occurrence of a Fano-type interference at that frequency, that arises from the interference between the localized, resonant, block modes, that are present at the shorter-wavelength side of the Reststrahlen band, and the extended, non-resonant \cite{nonres}, void modes, that are present at the longer-wavelength side of the Reststrahlen band. However, the spectral region we focus on here is the frequency regime near the blue bound of the Reststrahlen band. This is the frequency regime, where the localized block-cavity modes are excited, yielding the strong electric field required to achieve a strong absorption. In other words, the SiC blocks behave as microantennas in this spectral region. 
\par
Now, let us compare the spectral reflection response of the entire pyramid array with that of the individual-blocks arrays. We can observe that the micropyramid array has a hybrid response borrowing characteristics from the reflection response of the different individual-block arrays. In particular, the reflection response of the micropyramid array for the spectral regions of the Fano interference and void resonances is dominated by the reflection response of the largest constituent block. On the other hand,  the respective reflection response for the spectral region where the localized cavity modes are excited mimics the near-zero reflection characteristics of the smaller-sized blocks. This essentially means, that the pyramid arrangement of the individual microantenna blocks facilitates an efficient cascaded coupling \cite{cascaded} of EM energy  to each individual microantenna blocks. In this manner, the entire micropyramid acts like an efficient broadband micro-resonator, causing the trapped EM energy to get absorbed by the SiC material over a broad frequency range. 
\par 
This effect can be  visualized in Fig. 4, where we depict with white streamlines the EM energy circulation for three selected characteristic free space wavelengths, - denoted as M1, M2 and M3-. These characteristic free-space wavelengths were previously designated in Fig. 2, in panels (a)-(c). At the free space wavelength denoted M1, which is  10.5 $\mu$m [see Fig. 2] we can identify clear vortices around the sides or corners in all the SiC blocks comprising the micropyramid. These vortices in the EM energy streamlines signify the existence of trapped cavity-type of EM modes\cite{vortexcavity} in all the micropyramid's SiC blocks, similar to the ones we observed in the case of the individual blocks on their own in  Fig. 3 [see panel (c) for the first block and panel (d) for the fourth block]. This means that the cavity-modes in each block of the pyramid, synergistically contribute toward an increased interaction between the EM field and the SiC matter in the entire pyramid that yields the strong absorption response we have observed in Fig. 2(b). Indeed, the cavity-mode behavior as evidenced by the vortices in the EM streamlines leads to strongly enhanced fields within the micropyramid that we show in Fig. 4(d). For a more clear view of these fields we also show the field values only within the micropyramid, with a saturated color-map, with red signifying an electric field intensity enhancement of ten or larger value [see Fig. 4(g)]. 
\par
On the other hand, at free-space wavelength denoted M3, which is 12.0 $\mu$m, the incident EM wave essentially just streams downwards, completely avoiding the pyramid [see Fig. 4(c)]. In this case, the electric field intensity inside the micropyramid is much weaker than that of the incident light and essentially near-zero for most of the micropyramid [see Figs.  4(f) and 4(i)]. As a result, the micropyramid does respond with a low reflection, albeit with little absorption, as the incident light does not interact with the SiC absorbing matter. 
\par
For completeness, we also depict the situation at the free-space wavelength M2, which is 11.2 $\mu$m. This represents the transition from the block-cavity resonant response [for the free space wavelengths in the spectrum around M1], where the micropyramid is a strong absorber, to the EM down-streaming via the voids [for the free space wavelengths in the spectrum around M3], where the micropyramid is quite transparent. At this transitional regime we neither observe strong vortices, nor a down-streaming around the pyramid [see Fig. 4(b)], and the electric field intensity enhancement within the micropyramid is moderate [see Figs. 4(e) and 4(h)]. This situation actually represents the Fano interference, between the localized type of modes, existing in the spectrum around M1, and the extended modes, existing in the spectrum around M3, that results in the pronounced reflection peak we have observed in Fig. 2.
 
\section{V. HARNESSING SUPERABSORPTION IN THE MICROPYRAMID ARRAY SYSTEM}    
\par   
The focus of our discussion, from here there on,  would be only within the spectral regime around free-space wavelength M1, where the block-cavity modes responsible for the absorption enhancement are excited. Thus far, we have discussed the broadband absorption enhancement capabilities for sparse micropyramid arrays, with a micropyramid separation of 10 $\mu$m. This means, that the cavity modes in each micropyramid, interact very weakly since they are spatially apart by about a wavelength. It would be therefore interesting to explore how the cavity modes are affected when they can interact more strongly, and whether such stronger interaction can yield a stronger absorption-enhancement phenomenon. 
\par
We therefore calculate the spectral reflection and absorption response of arrays with closer-spaced micropyramids.  We show the results in Fig. 5 for interspacings of 7.5 $\mu$m and 5.0 $\mu$m  with dashed-red and solid-blue lines respectively, along with the original micropyramid array for comparison [black dotted lines]. We observe in Fig. 5(a) that the moth-eye-like SiC micropyramid building blocks continue to possess the broad-spectrum reflectionless capability, even when placed very closely. In particular, the array of micropyramid interspacings of 5.0 $\mu$m, leaves only small, one-micron-wide, voids at the base of the pyramid for light to squeeze through; this is about one-tenth the impinging light's wavelength. Yet, impressively it still possesses a near-reflectionless spectral response between 10.3 and 11 $\mu$m. We attribute this to the efficient cascaded coupling to each of the micro-antenna cavity modes of the individual blocks.
\par
Accordingly, since almost-all impinging EM energy is injected into the micropyramid array system in all cases, the key to the respective absorption performance lies in the electric field re-connfiguration that results from the interaction between the cavity modes in the adjacent micropyramids. Since the power dissipation rate per volume is proportional to the electric field intensity [see eq. A3], we can infer that the higher the electric field intensity the better the absorption performance. However, we should not quickly deduce from this fact, that the integrated electric field intensity within the micropyramid, in the new closer-spaced arrays, needs to be higher in comparison with the one of the original micropyramid array. It only needs to be higher than the ratio of the micropyramid interspacing, a, in the newly considered arrays over the respective value of the original array. This is because, for example an array with half the micropyramid interspacing of the original array, has twice as much SiC material underneath the same illumination area in comparison to the original array. This can be clearly understood in the appendix, where we have derived an analytic expression that explicitly relates the absorption by the SiC micropyramid array with the integrated normalized electric field intensity within the SiC material and the array's micropyramid interspacing, a (see eq. A.6 in appendix).
\par
In order to get a feeling for the electric field reconnfiguration emanating from the strong coupling between the adjacent cavity modes,  we plot the ratio between the electric field intensity in each location in the vicinity of the SiC micropyramid in the new arrays and the respective value in the original micropyramid array. The results are shown in Figs. 5(c) and (d) for the respective cases of  micropyramid interspacings of $\textrm a$=7.5 $\mu$m and $\textrm a$=5.0 $\mu$m. Indeed, we observe a strong electric-field reconnfiguration as a result of the strong coupling between the adjacent micropyramid modes. We observe, that there are locations where the electric field intensity is even enhanced; however in other areas we see that electric field intensity is weakened. If we now calculate the integrated value of the normalized electric field intensity, $E_{\textrm{norm}}$ \cite{norm}, at a free-space wavelength of 10.5 $\mu$m, over the extend of the SiC micropyramid structure, for both the new arrays, we find that:
\begin{equation}
\frac{\left. \int |E_{\textrm{norm}}|^2 dx dy \right|_{\textrm a=7.5 \mu \textrm m}}{\left. \int |E_{\textrm{norm}}|^2 dx dy \right|_{\textrm a=10 \mu \textrm m}}\sim 86.1 \%,
\end{equation} 
and\\     
\begin{equation}
\frac{\left. \int |E_{\textrm{norm}}|^2 dx dy \right|_{\textrm a=5 \mu \textrm m}}{\left. \int |E_{\textrm{norm}}|^2 dx dy \right|_{\textrm a=10 \mu \textrm m}}\sim 68.5\%.
\end{equation}
\par
From Eqs. (2) and (3) we expect that at the free-space wavelength of 10.5 $\mu$m, in the new arrays with $\textrm a$=7.5 $\mu$m and $\textrm a$=5.0 $\mu$m,  the absorption should be enhanced by respective factors of $\sim ~1.15$ and $\sim ~1.35$ \cite{factors} with respect to the absorption of the original array with $\textrm a$=10 $\mu$m. The expected absorption enhancement factors due to the electric field reconnfiguration, as quoted above, are in excellent agreement with the observed absorption enhancement that we observe in Fig. 5(b). 
\par
\section{VI. PERFORMANCE AND ROBUSTNESS OF THE MICROPYRAMID-ARRAY SUPERABSORBER}  
\par    
The closer-spaced micropyramid array, with interspacing of $\textrm a$=5.0 $\mu$m, demonstrates an extra-ordinary,-- more than 80$\%$ absorptance--, over a broad wavelength range between roughly 10.4 $\mu$m and 11 $\mu$m, thus behaving as a superabsorber \cite{superabsordef}. Note that such superabsorber behavior represents an extra-ordinary absorption enhancement in this range with respect to the absorption achieved by a bulk SiC slab that is about a wavelength thick [see Fig. 6(a)]. Such broadband superabsorber behavior is not affected by the presence of a transparent material substrate. We observe in Fig. 6(b) that the absorptance remains unaffected if the micropyramid array is placed on a 10 $\mu$m thick BaF$_2$ substrate (dotted-blue line in the figure). Note that BaF$_2$ is a fairly transparent material in the mid-IR wavelength range. For the calculations of Fig. 6(b) we have taken the refractive index of BaF$_2$ to be equal to 1.36.  
\par
Moreover, we find that the micropyramid design does not show sensitivity with respect to the alignment of the individual blocks. We show indicatively in Figs. 7(a) and 7(b) with a solid black line the results for the reflectance,   R, and the absorptance, A,  respectively, for the case of a misaligned-micropyramid array. For comparison the results for the symmetric micropyramid design are also depicted with dotted green lines. The off-center shift of the misaligned-pyramid design (see schematic in the top right panel of Fig. 7) is 150 nm, for the first block, 100 nm for the second block and 250 nm for the third block. We also show the EM energy circulation around the misaligned micropyramid in Fig. 7(c), along with the result for the corresponding symmetric design, i.e. the design without misalignment [seen in Fig. 7(d)]. 
\par
The depicted energy circulation is for wavelength M1, designated with the orange vertical line in the figure, for which cavity-localized modes are excited in the individual blocks. We observe in Fig. 7(c) the characteristic vortex-like EM circulation along the sides or corners of the individual blocks. These are similar to the ones seen in Fig. 7(d) for the corresponding symmetric micropyramid, or the modes we observed before in Fig. 4(a) (for the micropyramid array with interspacing a=10 $\mu$m). What is different here for the misaligned-pyramid design is that the mode field/energy landscape seizes to show the mirror symmetry with respect to the center axis of each individual block, which is the case for the $x$-axis symmetric pyramid designs of Figs. 4(a) and 7(d). In fact,  coupling to a certain side and/or corner appears stronger than the other. However, we see that the cascaded coupling from block to block is not obstructed. Therefore, the micropyramid still shows a low reflection/high absorption response in the wavelength regime around M1 (shaded green area in the figure), as we observe in Figs. 7(a) and 7(b).
\par
\section{VII. STRONGLY ASYMMETRIC RESPONSE IN THE SUPERABSORBER BEHAVIOR OF THE MICROPYRAMID ARRAY}   
\par
In all the results we have shown thus far, light is incident from the tip side of the micropyramid. It would be interesting to see what would happen if we reversed the micropyramid, so as light would be incident from the base side toward the tip of the micropyramid. We show the results for the reflection and absorption response of the reversely-oriented micropyramid in Figs. 8(a) and 8(b) respectively. It is impressive to observe that the broadband superabsorption behavior of the micropyramid strongly depends on the direction the light is coming from. In fact, we see that if we reverse the micropyramids in the array, absorption decreases dramatically [see red-dashed lines in Fig. 8(b)]. In other words, the micropyramid array is only a powerful absorber for light impinging from the pyramids' tip-side. 
\par
We stress that such interesting strongly asymmetric  response is not a mere outcome of the $y$-axis-asymmetric shape of the micropyramids. For example, an identical pyramid made from a non-lossy optical material would not exhibit any asymmetric reflection response at all. In such non-lossy structure, Lorentz reciprocity \cite{lorentz, vespe_book}, which mandates transmission, T, to be symmetric, i.e. the same both for tip-to-base and base-to-tip incidence, mandates also reflection, R to be symmetric. This is because in the absence of optical loss, R=1-T.  Now, in the lossy SiC micropyramid array we saw that reflection can be highly asymmetrical. However, this does not mean that  the system becomes non-reciprocal. Indeed, the system is Lorentz reciprocal with the transmission being the same for both tip-side and base-side incidence. Now, since R=1-T-A both reflection, R and absorptance, A, can become highly asymmetric while  transmission, T is symmetric. 
\par
In other words, material optical loss is necessary to obtain any asymmetry effects in reflection/absorption. Although necessary, optical material loss by itself is not sufficient to produce strongly asymmetric effects in reflection/absorption. The highly asymmetric behavior of the micropyramid array in the reflection and absorption is a result of the coupling to the cavity modes of the block encountered by the impinging light and  subsequent cascaded coupling to the cavity modes of the remaining SiC blocks [see Figs. 8(c) and 8(e)]. Note, that for the same SiC-micropyramid array at the frequency regime where the block cavity modes are not excited (i.e. outside the green shaded area in Fig. 8) [see Figs. 8(d) and 8(f)], the reflection and absorption are similar for both tip-side and base-side incidence. This stresses that a high dielectric loss factor in an asymmetrically shaped structure does not necessarily imply a strongly asymmetric reflection/absorption response. Our results indicate that the key protagonists to an asymmetric absorption/reflection response are cascaded resonances with highly asymmetric in-coupling. This new physical insight uncovered by our work establishes a transferable design principle towards achieving a strongly asymmetric reflection/absorption in other systems.
\par
 To our knowledge, such asymmetric reflection/absorption behavior of Fig. 8 in the frequency regime designated with the green shaded area is the first report of a strongly asymmetric reflection/absorption in passive systems with Reststrahlen-band materials \cite{Wassermanreview}. We note in passing that asymmetric absorption/reflection effects, albeit much weaker in comparison with the ones shown in Fig. 8, have been also reported with plasmonic systems in the visible range \cite{asym1, asym2, asym3}.  Also, although not originally studied within this context, plasmonic metasurface structures that are impedance matched with vacuum, connected to a ground plane via a dielectric spacer \cite{meta_asym1, meta_asym2, meta_asym3}, would be expected to respond with a highly asymmetric absorption/reflection. However, in this class of systems the plasmonic absorber material is separated by a dielectric spacer with a lower thermal conductivity which may limit their functionality as thermal emitters. In contrast, our proposed SiC micropyramid design involves a single-kind of connected absorber material of high thermal conductivity.
\par 
In the following section, we explore whether this observed asymmetric absorption response of the SiC micropyramid system can be further enhanced to achieve a near uni-directional absorption/emission behavior. Our objective will be to harness reflection to achieve as close as possible the following target: zero reflection for tip-to-base incidence and unity reflection for base-to-tip incidence, with transmission being zero for both incidences. It is important to have as close as possible to a zero transmission, i.e. an opaque behavior. Actually, opacity is a necessary condition to perfect emissivity, as a non-zero transmission makes a medium less emissive. Note, in the extreme case of a transparent system (i.e. with transmission one) emissivity is always zero \cite{Kirchhoff3}. A zero reflection for tip-to-base incidence and unity reflection for base-to-tip incidence, with opacity, implies a unity absorption for tip-to-base incidence and a zero absorption for base-to-tip incidence. Then Kirchhoff's law states \cite{Kirchhoff1, Kirchhoff2} that at a certain temperature, {\it T}, and wavelength, $\lambda$, the emissivity, {\it e}, of an opaque structure, towards a certain direction equals with the absorptance, A, for the same structure for light incident from that direction. In other words:\\
\begin{equation}
\begin{aligned}
e(\lambda, T;{\textrm{base-to-tip}}) &= \textrm A(\lambda, T;{\textrm{tip-to-base}})  &\\
 &=1-\textrm R(\lambda, T;{\textrm{tip-to-base}})
\end{aligned}
\end{equation}

\begin{equation}
\begin{aligned}
e(\lambda, T;{\textrm{tip-to-base}}) &=\textrm A(\lambda, T;{\textrm{base-to-tip}}) &\\
&=1-\textrm R(\lambda, T;{\textrm{base-to-tip}})
\end{aligned}
\end{equation}
Eqs. (4) and (5) make it clear why an opaque micropyramid system, with perfectly asymmetric reflection, zero and one respectively for base-to-tip and tip-to-base incidence, would emit only in the direction from the base to the tip of the micropyramid. In the following we investigate enhancing the observed asymmetry in absorption/reflection of the system of Fig. 8, as close as possible to the aforementioned perfect condition in order to enable a near-unidirectional emitter behavior for the micropyramid array.
\par
\section{VIII. NEAR UNI-DIRECTIONAL SUPERABSORBER/SUPER-EMITTER BEHAVIOR OF THE MICROPYRAMID ARRAY}
\par
Firstly, in order to strengthen the asymmetric reflection/absorption response of the micropyramid array, we explore increasing the thickness of the fourth block, while keeping the entire micropyramid thickness sub-wavelength (about half the free space wavelength). Therefore, we increase the thickness of the micropyramid's base block from 1 $\mu$m to 3 $\mu$m and refer to this modified micropyramid-array design as design B. In this section we focus only around the frequency regime where cavity modes are excited.   As we have discussed in Sec. VI this is the frequency region with the potential for a strongly asymmetric response. Fig. 9(a) depicts the spectral reflection response while Fig.  9(b) depicts the spectral absorption response of the design B micropyramid-array. The black-solid lines represent the result for tip-to-base incidence while the red-solid lines represent the result for base-to-tip incidence. The results of the original design of Fig. 8, which we will refer to as design A from thereon, are also included in Fig. 9 for comparison. The green-dotted lines designate the response of the original design A array for  tip-to-base incidence, while the blue-dotted lines designate the corresponding response for base-to-tip incidence. 
\par
We observe an increased reflection for the base-to-tip incidence for the modified design B array which results in a reduced absorption when compared with the original, design A, array. We attribute this to the larger size of the SiC block which comes at first contact with the impinging light for the base-to-tip incidence case. At the same time, for tip-to-base incidence we do not see any significant changes in the reflection response. This is because for such case, the first three blocks that the incident EM encounters are identical with the original, design A, array. However, the absorption ends-up being larger in the modified design B array  for tip-to-base incidence. This is because the same EM energy that gets sequentially coupled to the last block, now interacts with a larger volume of lossy matter. The combined effect of absorption decrease for base-to-tip incidence with absorption increase for tip-to-base incidence leads to a stronger asymmetry in the absorption response.  
\par
Secondly, we explore the effect of bringing the micropyramids closer. Therefore we start from the original design, design A, and bring each micropyramid building block closer at an interspacing of 4.2 $\mu$m. We refer to this modified design as design C.  We show the results for the reflection and absorption response in Figs. 9(c) and 9(d) respectively. Here also, the black-solid lines designate the results for tip-to-base incidence, while the red-solid lines designate the results for base-to-tip incidence. Like in the cases of Figs. 9(a) and 9(b), the original design A results are depicted as well for comparison purposes. Indeed, as expected this closer-spaced array shows a larger reflection and so a smaller absorption for base-to-tip incidence. This is because for such case the wave encounters thick closely-spaced SiC blocks resulting in a weaker coupling to the cavity modes. On the other hand, the tip-to-base incidence case is not affected much by bringing the micropyramids closer, as the wave still firstly encounters relatively sparsely-spaced small SiC blocks. Accordingly, since absorption does not change for tip-to-base incidence but decreases for base-to-tip incidence when compared with the original design A array, the net effect is that the asymmetry in the absorption response becomes stronger for the modified design C micro-array.  
\par
We then explore if both effects of asymmetry enhancement in the response of the two aforementioned designs can work in synergy when combined into one micropyramid-array design. We therefore explore a design where the micropyramids are more closely spaced at 4.2 $\mu$m, and have also their base block thicker (3 $\mu$m versus 1 $\mu$m of the original design of Fig. 8) [design BC]. We show the results for design BC in Fig. 10(a), for the reflection response and Fig. 10(b), for the absorption response. Black-solid and red-solid lines represent the results for tip-to-base and base-to-tip incidence respectively. Indeed, we find that the asymmetry in the absorption response gets further enhanced. 
\par
In particular, in the frequency region represented with the green shading in Fig. 10 we observe that the absorption is near-unity for tip-to-base incidence and very small (less than 20$\%$) for base-to-tip incidence. Essentially, the micropyramid array is an effective absorber only for light incident from the tip-side of the micropyramid. Moreover, in the green-shaded region the micropyramid system is near-opaque with transmission less than $2\%$. Accordingly, Eqs. (4) and (5) suggest that the micropyramid system of Fig. 10 would emit predominantly only in the direction from the micropyramid's base towards the tip. To be specific, the emission in the base-to-tip direction would be in the range of five to fifteen times stronger in the wavelength spectrum designated with the green shaded area in Fig. 10 with respect to the emission in the tip-to-base direction. In other words, the micropyramid array of Fig. 10 behaves as a near-unidirectional absorber/emitter. 
\par
\section{IX. CONCLUSIONS}   
\par   
We have presented a stepped micropyramid SiC array structure with super-absorber capabilities over a broad spectral range within the SiC phonon-polariton gap (Reststrahlen band). For TH-polarized light, this structure acts simultaneously as a moth-eye antireflector, allowing almost all light to couple inside, and a broad-band microantenna. The super-absorber capabilities emanate from the cascaded coupling of corner/side vortex modes at each of the SiC blocks comprising the micropyramid. The cascaded coupling of such modes from block-to-block of the lossy SiC micropyramid system along with the asymmetric in-coupling to these modes are the key protagonists that enable a highly uni-directional reflection/absorption response. Specifically, the micropyramid system is a strong absorber only for light impinging from the tips to the bases of the micropyramids.  These results and physical insight for the underpinning mechanisms provide transferable design principles towards achieving a strongly asymmetric broadband reflection/absorption in other systems. Furthermore, we discussed how the SiC micropyramid system, when near-opaque with a highly asymmetric reflection/absorption,  would behave as a near-unidirectional emitter by vitue of Kirchhoff's law \cite{Kirchhoff1, Kirchhoff2}.
\par
Our proposed platform could be applied in improving the efficiency and directionality of infrared globar-type of sources \cite{globar}. Moreover, the operational bandwidth of this system falls within the atmospheric transparency window \cite{atmosph, atmo}. At the same time SiC is a weakly absorbing material for most of the solar radiation spectrum. These two facts,  in combination, suggest that the micropyramid-array platform can be highly relevant to the emerging area of passive radiative cooling, a promising avenue for cooling buildings and  vehicles \cite{fan_rad, fan_rad2}. The highly unidirectional emission characteristics of our proposed platform may inspire designs for temperature management of electronic and plasmonic devices \cite{borisk1}, which is crucial to their resilience and functionality.  Moreover, our system studied here may inspire structures comprising other infrared active materials that could improve the sensitivity of infrared cameras and detectors \cite{detec1}.  
  
\section{Acknowledgements}   
Financial support for the Ph.D. studentship of G. C. R. Devarapu by the College of Engineering, Mathematics and Physical Sciences (CEMPS) University of Exeter is acknowledged.
\appendix
\numberwithin{equation}{section}
\renewcommand\theequation{\thesection A.\arabic{equation}} 
\section{Appendix: Absorptance and electric field intensity in the micro-pyramid array}
\setcounter{figure}{0}  
\setcounter{equation}{0}  
\par 
In the following, we derive an expression yielding the absorptance through the micropyramid array, A, versus the electric field intensity within the bounds of the micropyramid building block in each unit cell of the array. The total absorption through the stepped micropyramid array, should be the sum of the respective absorption,  A$(i)$, provided by each of the constituent blocks of the SiC pyramid structure, i.e: 
\begin{equation}  
\textrm A=\sum_{i=1}^4 \textrm A(i)
\end{equation}
The fraction of incident power that gets absorbed within the $ i^{th}$ SiC building block of the micropyramid, A$(i)$ can be obtained as:
\begin{equation}
\label{eq:a7.1power0}
A(i)=\dfrac{\overline{P_{loss, i}}}{\overline{P_{inc}}},
\end{equation} 
where $\overline{P_{loss, i}}$ represents the time-averaged power dissipation in the $ i^{th}$  SiC building block in the elementary unit cell of the micropyramid array. Conversely, 
$\overline{P_{inc}}$ represents the time-averaged power incident on the elementary unit cell of the micropyramid array. As there is translational symmetry along the $z$ direction, we consider the EM power that impinges through an area of $A_{inc}=\textrm a \cdot l_z$, with $l_z$ being the length of an arbitrary segment along the $z$-direction of the SiC block, and a, being the micropyramid's interspacing (see figure A.1).
\par
From Poynting's theorem  \cite{Jackson}, we have that the time-averaged power dissipation per unit volume of a material, $\overline{P_{loss, \text v}}$ is given by:
\begin{equation}
\label{eq:a7.1power1}
\overline{P_{loss,\text v}}=\dfrac{\omega\varepsilon_0\varepsilon_r''(\omega)}{2} |E|^2,
\end{equation} 
where $\varepsilon_0$ is the vacuum permittivity and $\varepsilon_r''(\omega)$ the imaginary part of the relative permittivity of the material at the frequency $\omega$ of the impinging wave.
Therefore, the power dissipated within the volume, $V_i=w_{xi} \cdot w_{yi} \cdot l_z$ in the $i^{th}$ SiC  micropyramid block in the elementary unit cell in the array,  $\overline{P_{loss, i}}$  will be:
       
\par 
\begin{equation}
\label{eq:a7.1power3}
\overline{P_{loss,i}}= \dfrac{\omega\varepsilon_0\varepsilon_r''l_z}{2} \int_{0}^{w_{x i}} \int_{0}^{w_{y i}} |E(x,y)|^2 \,dx dy,
\end{equation}
where the widths, $w_{x i}$, $w_{y i}$ are depicted in the schematics of  Fig. A.1. Note, that in the above integration the $(x,y)$ Cartesian coordinates have been off-set to be zero at the bottom left corner of the $i^{th}$ SiC block in the micropyramid.  

On the other hand, the time-averaged incident power   $\overline{P_{inc}}$ through the area, $A_{inc}$, depicted in the schematics of Fig. A.1 is:

\begin{equation}
\label{eq:a7.1power5}
{\overline{P_{inc}}}=\dfrac{\textrm a l_z}{2c\upmu_0}|E_{inc}|^2,
\end{equation} 
where $\upmu_0$ is the vacuum permeability and $c$ the vacuum speed of light.
Now with the use of Eqs. A.4 and A.5   Eq. A.2  we obtain:

\begin{equation}
\label{eq:a7.1power6}  
\medmuskip = 1mu    
\thinmuskip=1mu
\thickmuskip=1mu
{\textrm A}( i)= \frac{\varepsilon_r''(\omega) \varepsilon_0 \upmu_0 \omega c}{\textrm a |E_{inc}|^2}  \int_{0}^{w_{x i}} \int_{0}^{w_{y i}} |E(x,y)|^2 \,dx dy= \nonumber
\end{equation}
\begin{equation}
\frac{\varepsilon_r''\omega}{\textrm a c}\int_{0}^{w_{x i}} \int_{0}^{w_{y i}} |E_{norm}(x,y)|^2 \,dx dy,  
\end{equation}
where $|E_{norm}(x,y)|=\dfrac{|E(x,y)|}{|E_{inc}|}$ represents the electric field, within the SiC block, normalized by the incident electric field, occurring when an incident wave of frequency $\omega$ is incident on the structure. In other words, $|E_{norm}(x,y)|$ represents the electric field enhancement, at a certain location $(x,y)$ within the $i^{th}$ SiC block of the micropyramid. Eq. (A6) underlines the importance of obtaining an enhanced electric field in areas within the micropyramid (as we have seen in Fig. 4), in order to obtain an enhanced absorption. This in turn, stresses on the key role of the block-cavity modes on obtaining a strong absorption response. The total absorption through the entire micropyramid can be then obtained from the summation of the respective absorptions, $A(i)$, in each of the SiC blocks. 
\par   
 
\newpage
\begin{figure}   
\includegraphics[width=7.5cm]{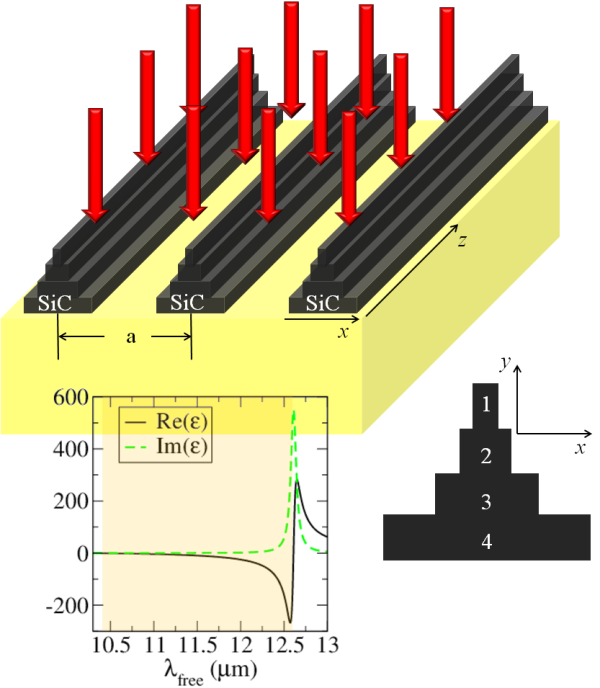}
\caption{The 2D SiC moth-eye superabsorber (the structure is assumed to be infinite in the $z$-direction). The bottom-right panel represents a front view of the  SiC stepped micropyramid, where we designate with a numeral each participating block. We also show in the bottom-left panel the permittivity, $\varepsilon$, of bulk SiC [real (black-solid line) and imaginary (dashed-green line) parts] in the Reststrahlen band (yellow shaded region), where the strong photon-phonon coupling yields a negative permittivity.}
\label{fig:1} 
\end{figure}
\begin{figure}
\includegraphics[width=7.5cm]{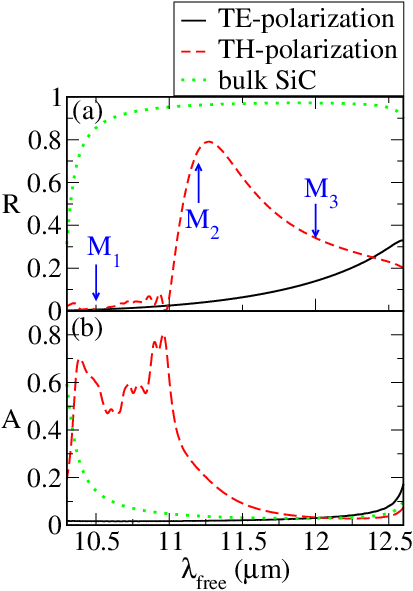}        
  
\caption{Spectral response versus the free space wavelength, $\lambda_{\textrm{free}}$ of the SiC-micropyramid array under normal incidence for the TE(TH) polarization case [solid-black (dashed-red) lines]. In (a) [(b)] the reflectance, R [absorptance, A] is shown. In addition, the reflection [absorption] from bulk SiC is shown in (a) [(b)] for comparison with dotted-green lines. We designate with $\textrm M_1$, $\textrm M_2$  and $\textrm M_3$ selected modes, each falling in one of the three characteristic spectral regimes: the broad-band reflectionless region, the region near the reflection maximum, and the second low-reflectivity window past the occurrence of the reflection peak.}
\end{figure}  
\par      
\begin{figure}[!htb]      
\includegraphics[width=7.5cm]{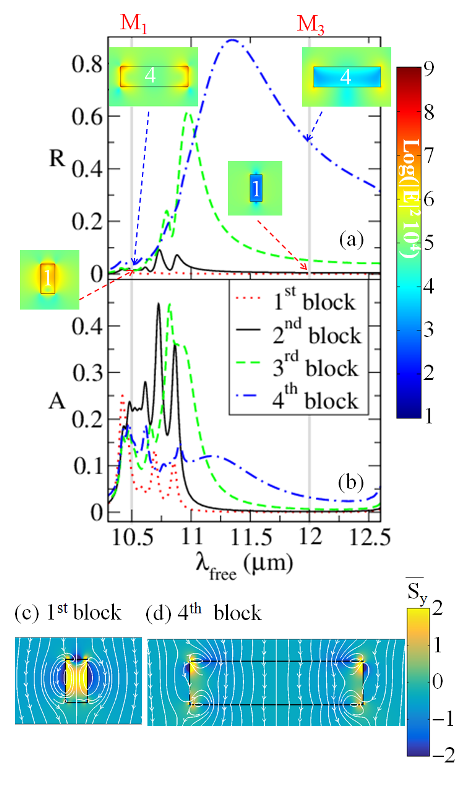}

\caption{Spectral reflection response [in (a)] and absorption response [in (b)] of the different periodic arrays made of the respective individual blocks comprising the pyramid of Fig. 1. The two characteristic frequencies/free space wavelengths, in the vicinity of the cavity and void resonances,  marked as M1 and M3 respectively, are designated with the gray vertical lines. The insets depict the electric field intensity around the SiC block \cite{norm}, for the respective arrays comprising the first or fourth of the micropyramid’s blocks for these two frequencies. Note, the corresponding color-map on the top right of the figure is in logarithmic scale. In addition, we show in panels (c) and (d) the energy circulation (white streamlines with arrows) around the first and fourth block respectively for the frequency M1. Note in (c) and (d), the background color-map represents the y-component of the time-averaged Poynting vector, $\bar S_y$ \cite{units}; hence negative values represent a downward EM flow.}  
\end{figure}

\par  
\begin{figure*}[!htb]        
\includegraphics[width=13.5cm]{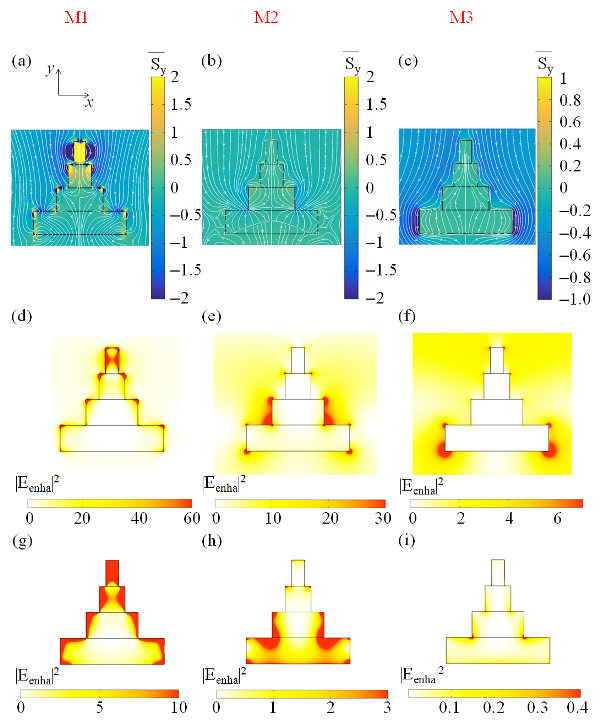}    
\caption{(a)-(c) Energy circulation (white streamlines with arrows) around the micropyramid array building blocks for three characteristic free-space wavelengths, designated in the corresponding spectral reflectance response of Fig. 2; M1: coupling to cascaded cavity resonances in the SiC blocks, M2: Fano interference M3: EM energy streams downwards through the voids. Note, the background color-map represents the y-component of the time-averaged Poynting vector, {\bf S} \cite{units}; hence negative values represent a downward EM flow. (d)-(f) Corresponding electric-field intensity enhancement (g)-(i) Same as in (d)-(f) but only the electric-field intensity enhancement inside the micropyramid is shown with a saturated colormap [red is used for any electric-field intensity enhancement that is higher than the maximum value of the associated colormap]}
\end{figure*}        
\par   
\par   
\begin{figure}[!htb]      
\includegraphics[width=7.5cm]{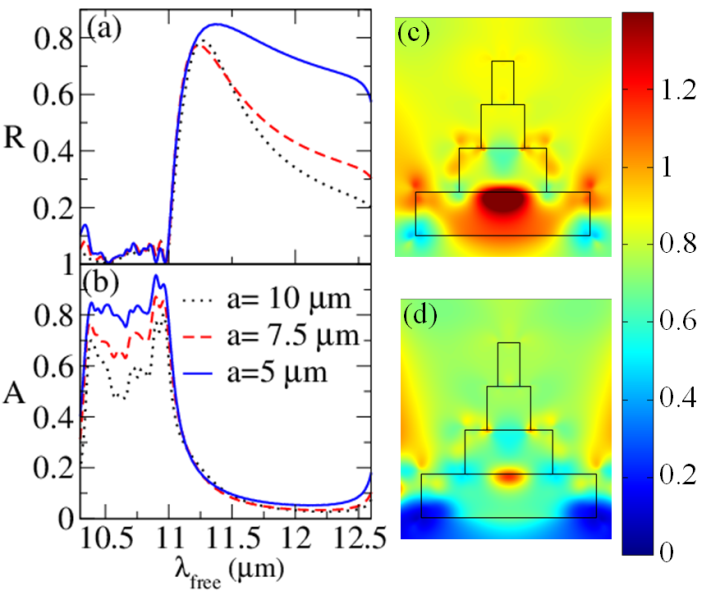}  
\caption{(a) Spectral reflectance, R, for the new arrays with micropyramid interspacings, a=7.5 $\mu$m (dashed-red lines) and a=5 $\mu$m (solid-blue lines). The reflectance of the original array with interspacing a=10 $\mu$m (dotted-black lines) is also shown for comparison. (b) Same as in (a) but for the absorptance, A. (c) Electric field intensity for the array with  interspacing, a=7.5 $\mu$m, with respect to the corresponding values of the original micropyramid array with a= 10 $\mu$m. (d) Same as in (c) but for the array with  interspacing, a=5 $\mu$m.}  
\end{figure}        
\par 
\par  
\begin{figure}[!htb]             
\includegraphics[width=7.5cm]{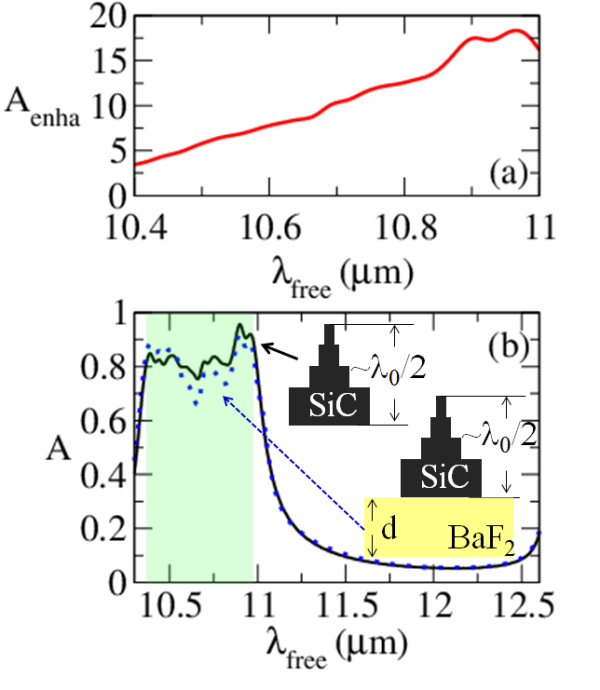}    
\caption{(a) Absorption enhancement, $\textrm A_{\textrm{enha}}$, exhibited by the micropyramid-array superabsorber of interspacing a=5 $\mu$m. (b)The influence of a $\textrm{BaF}_{\textrm 2}$ substrate of thickness d=10 $\mu$m,  on the absorptance, A, is shown with a dotted-blue line. For comparison, the absorptance for the array without the substrate is shown with a solid-black line. The green shaded area designates the spectrum where the  micropyramid-array with interspacing a=5 $\mu$m behaves as a superabsorber \cite{superabsordef}.}  
\end{figure}      
\par
\begin{figure}[!htb]            
\includegraphics[width=7.5cm]{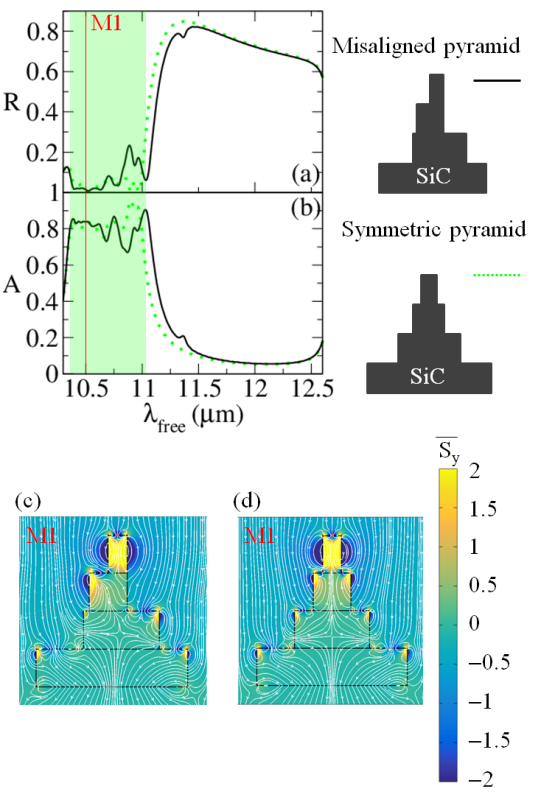}    
\caption{ Spectral response of reflectance [in (a)] and absorptance [in (b)]. Results are shown with a solid black for the case of a misaligned-pyramid array [see schematic depicted at the right of fig. (a)]. The corresponding results for the symmetric micropyramid design [depicted at the right of fig. (b)] are also shown with a dotted-green line for comparison. The shaded green area represents the spectrum where localized cavity resonances are excited in the individual blocks. The corresponding EM energy circulation for wavelength M1, designated with the vertical line in (a)-(b),  is shown in (c) for the misaligned-pyramid array and in (d) for the corresponding symmetric one. Note that as in Figs. 4(a)-(c), the background in (c) and (d) represents the $y$-component of the time-averaged Poynting vector \cite{units}. }  
\end{figure}      
\par  
\par    
\begin{figure}[!htb]              
\includegraphics[width=7.5cm]{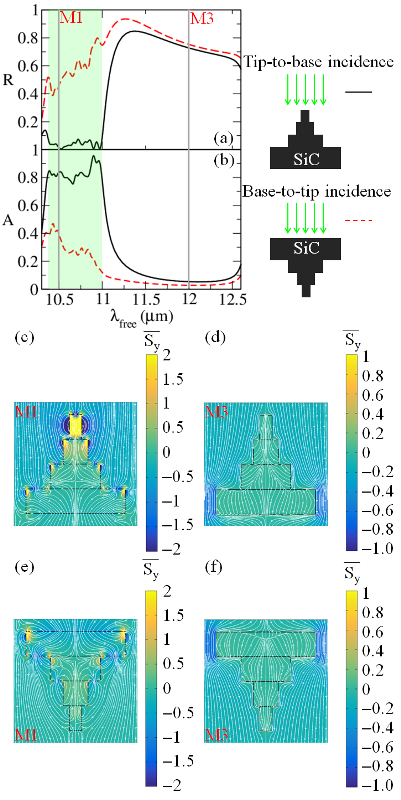}       
\caption{  Strongly asymmetric reflection/absorption response of the micropyramid array system with a= 5$\mu$m. The reflection, R, and absorption, A, for EM waves incident from the tip side of the micropyramid are shown with solid black lines in panels (a) and (b) respectively. Conversely, R, and A, are shown with dashed-red lines in the same figures, for EM waves incident from the base side of the micropyramid a. The green shaded area designates the spectral regime with a strong asymmetry in the reflection/absorption response between the tip-to-base and base-to-tip incidence. (c)-(d) EM energy circulation for wavelengths M1 and M3 for the case of incidence from the tip side of the micropyramid (e)-(f) EM energy circulation for wavelengths M1 and M3 for the case of incidence from the base side of the micropyramid. The wavelengths M1 and M3 are designated in (a)-(b) with gray vertical lines. Note that as in Figs. 4(a)-(c), the background in (c) through (f) represents the $y$-component of the time-averaged Poynting vector \cite{units}. }
\end{figure} 
\par
\par
\begin{figure}[!htb]               
\includegraphics[width=7.5cm]{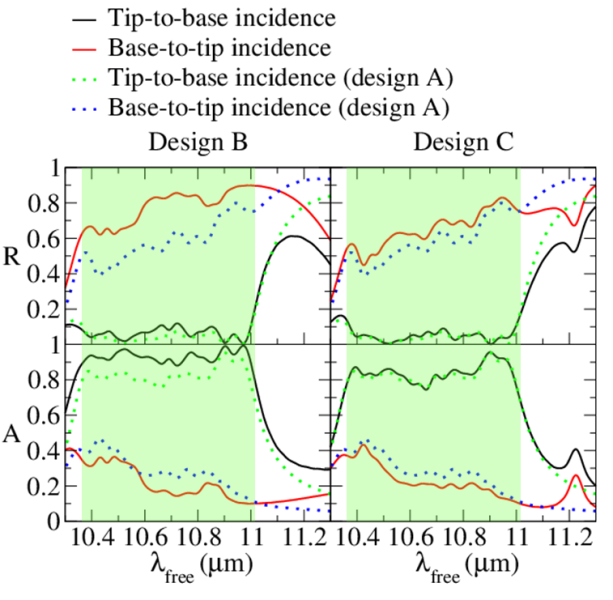}    
\caption{ Enhancing the asymmetry in the reflection/absorption response between tip-to-base and base-to-tip incidence by considering a thicker, 3 $\mu m$-thick  block for the pyramid base (design B), or by bringing the micropyramids closer at an interspacing of 4.2 $\mu m$ (design C). The reflectance, R, (top panels) and absorptance, A (bottom panels) are shown for each design with black-solid lines (red-solid lines) for tip-to-base (base-to-tip) incidence. For comparison the respective results for the design of Fig. 8 (design A) are shown with green-dotted lines (tip-to-base incidence) and blue-dotted-lines (base-to-tip incidence). The green shading designates the frequency region of strong asymmetry in the reflection/absorption response between tip-to-base incidence and base-to-tip incidence}  
\end{figure}       
\par
\begin{figure}[!htb]               
\includegraphics[width=7.5cm]{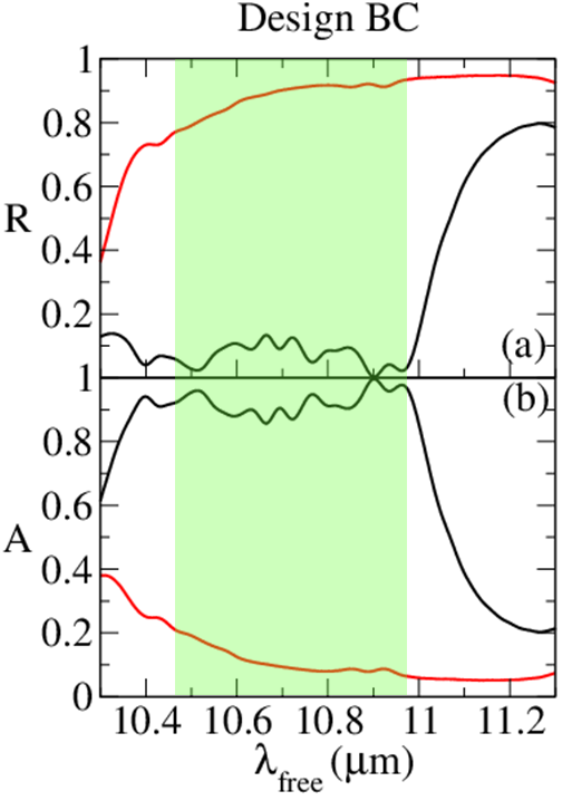}    
\caption{ The micropyramid as a near-unidirectional absorber/emitter. Reflection, R, [(a)] and absorption, A, [(b)] versus free space wavelength are shown for a micropyramid array design (design BC) with interspacing a=4.2 $\mu$m and micropyramid block sizes same as the design of Figs. 2/4 except for the fourth block being thicker (3 $\mu$m thick). The green-shading designates a frequency regime where the micropyramid absorbs with near $100\%$ efficiency when light incident from the tip-side of the micropyramid while it absorbs weakly, less than $20\%$, when light is incident from the base-side of the micropyramid in this manner acting as a uni-directional absorber/emitter.}  
\end{figure}      
\par
\renewcommand{\thefigure}{A.1} 
\begin{figure}[!htb]            
\includegraphics[width=7.5cm]{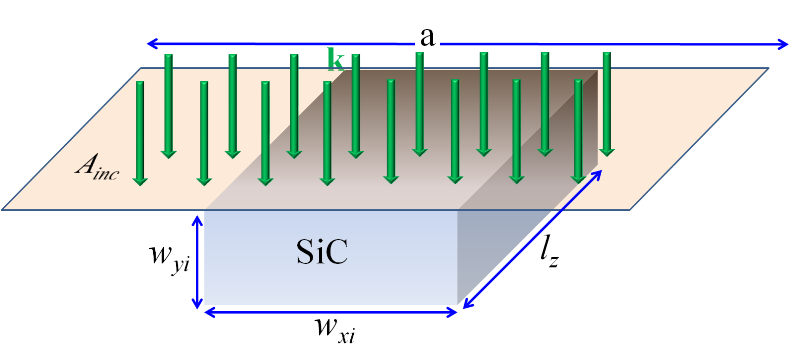}    
\caption{ EM radiation impinging onto the $i^{th}$ SiC block of the micropyramid in the elementary unit cell of the array with periodicity, a. The widths in the {\it x} and {\it y} direction of the block are designated. Due to translation symmetry in the {\it z}-direction, the segment with length $l_z$ is chosen arbitrarily. } 
\end{figure}


\begin{thebibliography}{99}    
\item[*]Corresponding author e-mail: sfoteino@unm.edu 
\bibitem{atwater} H. A. Atwater and A. Polman, Plasmonics for improved photovoltaic devices, Nat. Mat. {\bf 9}, 205 (2010).
\bibitem{aydin} K. Aydin, V. E. Ferry, R. M. Briggs, and H. A. Atwater, Broadband, polarization-independent resonant light absorption using ultrathin, plasmonic super absorbers,
Nat. Commun. {\bf 2}, 517 (2011).  
\bibitem{brongersma} R. A Pala, J. White, E. Barnard, J. Liu, and M. L. Brongersma, Design of plasmonic thin‐film solar cells with broadband absorption enhancements, Adv. Mat. {\bf 21}, 3504 (2009).
\bibitem{povinell} C. Lin, and M. L. Povinelli, Optical absorption enhancement in silicon nanowire arrays with a large lattice constant for photovoltaic applications ,
Opt. Express 17, 19371 (2009).
\bibitem{padilla1} C. M. Watts, X. Liu, and W. J. Padilla, Metamaterial electromagnetic wave absorbers, Adv. Mat. {\bf 24}, OP98 (2012). 
\bibitem{detec1} A. V. Barve, S. J. Lee, S. K. Noh, and S. Krishna, Review of current progress in quantum dot infrared photodetectors, Laser and Phot. Rev. {\bf 4}, 738 (2010).
\bibitem{wasser1} J. A. Mason, S. Smith, and D. Wasserman,  Strong absorption and selective thermal emission from a midinfrared
metamaterial, Appl. Phys. Lett. {\bf 98}, 241105 (2011).
\bibitem{detec2} Y. D. Sharma, Y. C. Jun, J. O. Kim, I. Brener and S. Krishna, Polarization-dependent photocurrent enhancement in metamaterial-coupled quantum dots-in-a-well infrared detector, Opt. Com. {\bf 312}, 31 (2014).
\bibitem{kats1} M. A. Kats, D. Sharma, J. Lin, P. Genevet, R. Blanchard, Z. Yang, M. M. Qazilbash, D. N. Basov, S. Ramanathan and F. Capasso, Ultra-thin perfect absorber employing a tunable phase change material, Appl. Phys. Lett. {\bf 101}, 221101 (2012).
\bibitem{modul} D. Shrekenhamer, J. Montoya, S. Krishna, and W. J. Padilla, Four-color Metamaterial Absorber THz Spatial Light Modulator, Adv. Opt. Mat. {\bf 1}, 905 (2013).
\bibitem{wasser2} W. Streyer, S. Law, A. Rosenberg, C. Roberts, V. A. Podolskiy, A. J. Hoffman, and D. Wasserman, Engineering absorption and blackbody radiation in the far-infrared with surface phonon polaritons on gallium phosphide, Appl. Phys. Lett. {\bf 104}, 131105 (2014).
\bibitem{shvets1} B. Neuner III, C. Wu, G. T. Eyck, M. Sinclair, I. Brener, and G. Shvets,  Efficient infrared thermal emitters based on low-albedo polaritonic meta-surfaces, Appl. Phys. Lett. {\bf 102}, 211111 (2013).
\bibitem{durdu1} A. Vora,  J. Gwamuri,  J. M. Pearce,  P. L. Bergstrom, and  D. O. Guney, Multi-resonant silver nano-disk patterned thin film hydrogenated amorphous silicon solar cells for Staebler-Wronski effect compensation, J. Appl. Phys. {\bf 116}, 093103 (2014).
\bibitem{shalaevbook} W. Cai and V. Shalaev, {\it Optical Metamaterials Fundamentals and Applications}, Springer, New York, (2010).
\bibitem{soukreview} C. M. Soukoulis and M. Wegener, Past achievements and future challenges in the development of three-dimensional photonic metamaterials, Nat. Photonics {\bf 5}, 523 (2011).
\bibitem{padilla2} N. I. Landy, S. Sajuyigbe , J. J. Mock, D. R. Smith and W. J. Padilla, Perfect Metamaterial Absorber, Phys. Rev. Lett. {\bf 100}, 207402 (2008).
\bibitem{metaabsor} A. Balmakou, M. Podalov, S. Khakhomov, D. Stavenga, and I. Semchenko, Ground-plane-less bidirectional terahertz absorber based on omega resonators,  Opt. Lett. {\bf 40}, 2084-2087 (2015).
\bibitem{boltasseva} J. J. Liu, G. V. Naik, S. Ishi, C. DeVault, A. Boltasseva, V. M. Shalaev, and E. Narimanov, Optical absorption of hyperbolic metamaterial with stochastic surfaces , Opt. Express {\bf 22}, 8893 (2014).
\bibitem{kats2} M. A. Kats, R. Blanchard, P. Genevet and F. Capasso, Nanometre optical coatings based on strong interference effects in highly absorbing media, Nature Materials {\bf 12}, 20 (2013).
\bibitem{sf1} G. C. R. Devarapu and S. Foteinopoulou, Mid-IR near-perfect absorption with a SiC photonic crystal with angle-controlled polarization selectivity, Opt. Express {\bf 20}, 13041  (2012).
\bibitem{sf2} G. C. R. Devarapu and S. Foteinopoulou, Compact photonic-crystal superabsorbers from strongly absorbing media, J. Appl. Phys. {\bf 114}, 033504 (2013). 
\bibitem{photobook1} J. L. Gray, {\it The Physics of Solar Cells}, in {\it Handbook of Photovoltaic Science and Engineering}, ed. A. Luque and S. Hegedus, Wiley (2010).
\bibitem{biswas} B. Curtin, R. Biswas and V. Dalal, Photonic crystal based back reflectors for light management and enhanced absorption in amorphous silicon solar cells, Appl. Phys. Lett. {\bf 95}, 231102 (2009). 
\bibitem{hessel_rev} B. Lee, Il-Min Lee, S. Kima, D.-H. Oh and L. Hesselink, Review on subwavelength confinement of light with plasmonics, J. of Mod. Optics {\bf 57}, 1479 (2010).
\bibitem{catrysse} P. B. Catrysse and S. Fan, Near-complete transmission through subwavelength hole arrays in phonon-polaritonic thin films, Phys. Rev. B {\bf 75}, 075422 (2007).
\bibitem{biosens} A. Ganjoo, H. Jain, C. Yu, J. Irudayaraj, and C. G. Pantano, Detection and fingerprinting of pathogens: Mid-IR
biosensor using amorphous chalcogenide films, J. Non-Crystalline Solids {\bf 354}, 2757 (2008).
\bibitem{fan_rad} A. P. Raman, M. A. Anoma, L. Zhu, E. Rephaeli and S. Fan, Passive radiative cooling below ambient air temperature under direct sunlight, Nature {\bf 515}, 540 (2014).
\bibitem{atmosph} C. Kuenzer, S. Dech, {\it Theoretical background of thermal infrared remote sensing}, in {\it Thermal Infrared Remote sensing: Sensors, Methods and Applications}, ed. by C. Kuenzer, S. Dech, Springer (2013).  
\bibitem{atmo} M. Cucumo, A. De Rosa, V. Marinelli, Experimental testing of correlations to calculate the atmospheric “transparency window” emissivity coefficient, Solar Energy {\bf 80}, 1031 (2006).
\bibitem{shalaev_refra} U. Guler, A. Boltasseva and V. M. Shalaev, Refractory Plasmonics, Science {\bf 344}, 263 (2004).  
\bibitem{superabsordef} With the term {\it superabsorber} we imply a system demonstrating a strong absorption, that can reach more than 80$\%$, which at the same time represents a strong absorption enhancement with respect to the absorption exhibited by the lossy constituent material, SiC in our case, when in bulk form. Please see also references to superabsorber behavior exhibited in other systems \cite{aydin, superabsordef2, superabsordef3}
\bibitem{superabsordef2} N. Zhang, K. Liu, H. Song, Z. Liu, D. Ji, X. Zeng, S. Jiang, and Q. Gan, Appl. Phys. Lett.{\bf 104}, 203112 (2014).
\bibitem{superabsordef3} V. Yannopapas, and I. E. Psarobas, Ordered Arrays of Metal Nanostrings as Broadband Super Absorbers, J. Phys. Chem. C {\bf 116}, 15599 (2012).
\bibitem{Kirchhoff1} F. E. Nicodemus, Directional Reflectance and Emissivity of an Opaque Surface, Applied Opt. 4, 767 (1965).
\bibitem{Kirchhoff2} Eugene A. Sharkov, Passive Microwave sensing of the Earth, Springer (2003).
\bibitem{jeosrp} G. C. R. Devarapu and S. Foteinopoulou, Broadband Mid-IR superabsorption with aperiodic polaritonic photonic crystals, J. Eur. Opt. Soc.:Rapid Pub. {\bf 9}, 14012 (2014).   
\bibitem{moth1} S. J. Wilson and M. C. Hutley, The Optical properties of ``moth-eye'' antireflection surfaces, Opt. Acta {\bf 29}, (1982).  
\bibitem{moth2} P. Kunze and K. Hausen, Inhomogeneous refractive index in the crystalline cone of a moth eye, Nature {\bf 231}, 392 (1971).
\bibitem{rivas} S. L. Diedenhofen, G. Vecchi, R. E. Alga, A. Hartsuiker, O. L. Muskens, G. Immink, E. P. A. M. Bakkers, W. L. Vos, and J. G. Rivas, Broad-band and omnidirectional antireflection coatings based on semiconductor nanorods, Adv. Mat. {\bf 21}, 973 (2009). 
\bibitem{biomime2} W.-L. Min, B. Jiang and P. Jiang, Bioinspired self‐cleaning antireflection coatings, Adv. Mat. {\bf 20}, 3914 (2008).  
\bibitem{pwe} P. R. Villeneuve and M. Piche, Photonic bandgaps in periodic dielectric structures, Prof. Quant. Elec. {\bf 18}, 153 (1994).
\bibitem{SiCwafer} G. N. Yushin, A. V. Kvit, R. Collazo, and Z. Sitar, SiC to SiC wafer bonding, MRS Proc. {\bf 742}, (2002).
\bibitem{gsubra_ebeam} G. Subramania and S. Y. Lin, Fabrication of three-dimensional photonic crystal with alignment based on electron
beam lithography, Appl. Phys. Lett. {\bf 85}, 5037 (2004). 
\bibitem{SiCetch} J. Y. Lee, X. Lu, and Q. Lin, High-Q silicon carbide photonic-crystal cavities, Appl. Phys. Lett. {\bf 106}, 041106 (2015). 
\bibitem{bronger2} J. A. Schuller, T. Taubner, and M. L. Brongersma, Nat. Photonics {\bf 3}, 658 (2009).
\bibitem{lumerical} Lumerical FDTD Solutions; Reference Guide for FDTD Solutions http://www.lumerical.com/tcad-products/fdtd/
\bibitem{norm} The electric field intensity is normalized by the electric field intensity of the impinging EM wave, $|E_0|^2$ 
\bibitem{units} ${\bar{S_y}}=-\frac{1}{2} Re(E_x H_z*)$ is represented in units of $S_0=\frac{1}{c \mu_0} |E_0|^2$ with $|E_0|$ being the electric field amplitude of the source.
\bibitem{vortexcavity} Y.-L. Ho, L.-C. Huang, E. Lebrasseur, Y. Mita, and J.-J. Delaunay, Independent light-trapping cavity for ultra-sensitive plasmonic sensing, Appl. Phys. Lett. {\bf 105}, 061112 (2014).
\bibitem{fanoyuri} A. E. Miroshnichenko, S. Flach, and Y. S. Kivshar, Fano resonances in nanoscale structures 
, Rev. Mod. Phys. {\bf 82}, 2257 (2010).
\bibitem{Wassermanreview} K. Feng, W. Streyer, Y. Zhong, A.J. Hoffman, and D. Wasserman, Photonic materials, structures and devices for Reststrahlen optics,  Opt. Express {\bf 23},
A1418 (2015).
\bibitem{nonres} G. Subramania, S. Foteinopoulou and I. Brener, Nonresonant Broadband Funneling of Light via Ultrasubwavelength Channels, Phys. Rev. Lett. {\bf 107}, 163902 (2011).  
\bibitem{cascaded} R. Garg, K. Thyagarajan, Cascaded coupling: Realization and application to spectral maneuvering, Opt. Fiber Tech. {\bf 19}, 148 (2013).
\bibitem{factors} To obtain the absorption enhancement factors we divide the values of Eqs. (2) and (3) with the respective ratios of the micropyramid interspacing, a. 
\bibitem{lorentz} H. A. Lorentz, The theorem of Poynting concerning the energy in the electromagnetic field and two general propositions concerning the propagation of
light, Afd. Natuurkd. K. Ned. Akad. 4, 176 (1896).
\bibitem{vespe_book} M. Nieto-Vesperinas, {\it Scattering and Diffraction in Physical Optics}, Wiley, New York, (1991).
\bibitem{asym1} E. Altewischer, M. P. van Exter, and J. P. Woerdman, Nonreciprocal reflection of a subwavelength hole array , Opt. Lett. {\bf 28}, 1906 (2003).
\bibitem{asym2} S. Butun and K. Aydin, Asymmetric Light Absorption and Reflection in Freestanding Nanostructured Metallic Membranes, ACS Phot. {\bf 2}, 1652 (2015).
\bibitem{asym3} Note, the authors in Ref. \onlinecite{asym1} use the term ``non-reciprocal'' in the context of asymmetric reflection without violation of the Lorentz reciprocity principle.
\bibitem{meta_asym1} Xianliang Liu, Tatiana Starr, Anthony F. Starr, and Willie J. Padilla, Infrared Spatial and Frequency Selective Metamaterial with Near-Unity Absorbance, Phys. Rev. Lett. 104, 207403 (2010).
\bibitem{meta_asym2} C. Wu and G. Shvets,  Design of metamaterial surfaces with broadband absorbance, Opt. Lett. {\bf 37}, 308 (2012). 
\bibitem{meta_asym3} A. Vora, J. Gwamuri, N. Pala, A. Kulkarni, J. M. Pearce, and D. O. Guney, Exchanging Ohmic Losses in Metamaterial Absorbers with Useful Optical Absorption for Photovoltaics, Sci. Rep. {\bf 4}, 4901 (2014).
\bibitem{Kirchhoff3} D. P. DeWitt and F. P. Incroprera, in Theory and Practice of Radiation Thermometry, Chapter 1: Physics of Thermal Radiation, Wiley Interscience (1988).
\bibitem{globar} J. E. Stewart and J. C. Richmond, Infrared emission spectrum of silicon carbide heating elements, J. of Research of the National Bureau of Standards {\bf 59}, 2810 (1957).
\bibitem{fan_rad2} E. Raphaelli, A. Raman, and S. H. Fan, Ultrabroadband photonic structures to achieve high-performance
daytime radiative cooling, Nano Letters {\bf 13}, 1457 (2013).
\bibitem{Jackson} J. D. Jackson, {\it Classical Electrodynamics} (Third edition, John Wiley and Sons, Hoboken, 1998).
\bibitem{borisk1} S. V. Boriskina, J. K. Tong, W.-C. Hsu, L. Weinstein, X. Huang, J. Loomis, Y. Xu, G. Chen,  Hybrid optical-thermal devices and materials for light manipulation and radiative cooling, Proc. SPIE 9546, Active Photonic Materials VII, 95461U (2015).
\bibitem{borisk2}  S. V. Boriskina, J. K. Tong, W.-C. Hsu, B. Liao, Y. Huang, V. Chiloyan, G. Chen, {\it Heat meets light on the nanoscale}, Nanophotonics 5, 134 (2016).
\end{thebibliography}
\end{document}